\documentclass[aps,pra,twocolumn,letterpaper,10pt,superscriptaddress]{revtex4-2}

\usepackage{amssymb,amsthm,amsmath,amsfonts,mathcmd,mathrsfs}
\usepackage{graphicx,ulem,enumerate,bm,bbm,mathptmx}

\usepackage[pdftex,dvipsnames,usenames]{xcolor}
\usepackage[colorlinks=true,urlcolor=blue,citecolor=blue,linkcolor=blue]{hyperref}

\usepackage{braket,tensor}
\usepackage{stmaryrd,tikz}
\usepackage{esint}

\begin{document}

\title{Restricted Monte Carlo wave function method and Lindblad equation\\{}for identifying entangling open-quantum-system dynamics}

\author{Laura Ares}
    \affiliation{Institute for Photonic Quantum Systems (PhoQS), Paderborn University, Warburger Stra\ss{}e 100, 33098 Paderborn, Germany}

\author{Julien Pinske}
    \affiliation{Niels Bohr Institute, University of Copenhagen, Blegdamsvej 17, DK-2100 Copenhagen, Denmark}

\author{Benjamin Hinrichs}
    \affiliation{Institute for Photonic Quantum Systems (PhoQS), Paderborn University, Warburger Stra\ss{}e 100, 33098 Paderborn, Germany}

\author{Martin Kolb}
    \affiliation{Institute for Photonic Quantum Systems (PhoQS), Paderborn University, Warburger Stra\ss{}e 100, 33098 Paderborn, Germany}

\author{Jan Sperling}
    \affiliation{Institute for Photonic Quantum Systems (PhoQS), Paderborn University, Warburger Stra\ss{}e 100, 33098 Paderborn, Germany}

\date{\today}

\begin{abstract}
    We develop an extension of the Monte Carlo wave function approach that unambiguously identifies dynamical entanglement in general composite, open systems.
    Our algorithm performs tangential projections onto the set of separable states, leading to classically correlated quantum trajectories.
    By comparing this restricted evolution with the unrestricted one, we can characterize the entangling capabilities of quantum channels without making use of input-output relations.
    Moreover, applying this method is equivalent to solving the nonlinear master equation in Lindblad form introduced in \cite{PAH24} for two-qubit systems.
    Our approach also provides a necessary and sufficient condition for entanglement in multipartite systems of arbitrary size, in terms of a stochastic differential equation.
    Thereby, we isolate the impact of dynamical entanglement in open systems by applying our approach to several correlated decay processes.
    Therefore, our methodology provides a complete and ready-to-use framework to characterize dynamical quantum correlations caused by arbitrary open-system processes.
\end{abstract}

\maketitle

\section{Introduction}

    The superposition principle is an indispensable component of quantum mechanics whose consequences challenge our intuition about the behavior of nature \cite{D47}.
    Among these consequences, quantum entanglement has acquired an outstanding role among the various quantum correlations in composite quantum systems \cite{HH09}.
    It has emerged as the fundamental ingredient for quantum technologies, specially for quantum communication \cite{BS99,B02,HD08,BPMEWZ09,HGLLG23} and quantum computing \cite{RB01,RB02,G96,HD07}.

    While certifying entanglement is an NP-hard problem \cite{G03,I07}, many approaches to detect the presence of quantum entanglement have been proposed \cite{HH96,HH00,T00}.
    A highly effective strategy in this regard is the construction of entanglement witnesses \cite{GT09,LK00,HE06}. 
    One particular method to construct optimal witnesses is via the so-called separability eigenvalue equations \cite{SV09,SV13}, which was used to experimentally identify path-entangled photons \cite{GS16}, complex multipartite entanglement in experiments with frequency combs \cite{GS15}, and theoretical studies of entanglement in macroscopic systems \cite{SW17(b)}.

    Like all witnessing approaches, the separability eigenvalue method is restricted to stationary scenarios.
    Yet, this can be overcome by deriving a separability Schr\"odinger-type equations \cite{SW17} that capture the dynamical aspects of entanglement as well.
    These equations highlight temporal quantum effects of the inseparable evolution by comparing the trajectories to their classically correlated counterparts \cite{SW20}, thereby identifying inseparable and time-dependent quantum properties.

    Still, the notion of entangling dynamics extends far beyond unitary evolution;
    see, e.g., Refs. \cite{WB06,I09,GG12}.
    The most general propagation of a quantum state is given by the formalism of dynamical (completely positive) maps \cite{K83}, encompassing processes such as non-deterministic state preparation, noisy propagation, measurements, etc.
    Dynamical maps that describe a noisy evolution can, under certain mathematical conditions, be obtained as the solution of a quantum master equation in Lindblad form \cite{L76,GK76}.
    The set of processes which can be accurately described by a Lindblad master equation became more inclusive in recent years \cite{NR20,BS22}.
    In particular, the study of Markovian and non-Markovian master equations led to an extensive literature;
    see, e.g., Refs. \cite{BP02,BL16,RH12} and the references therein.

    In many cases, the solution of a quantum master equation has to be determined numerically \cite{DCM92,DPZG92,HW92,C93}.
    In particular, the Monte Carlo wave function approach \cite{MCD93,PK98} offers a computational advantage in extracting open-system dynamics by averaging all the possible pure-state trajectories resulting from random quantum jumps.
    This approach, being equivalent to solving the Lindblad equation, is generally applicable and has been applied in a wide variety of physical scenarios \cite{KLR22,GG23,KK20,HOR23}.

    Despite the success of all aforementioned approaches, what is missing to date is a framework that renders it possible to identify the temporal aspects of entanglement in processes that are governed by open-system features.
    Moreover, to quantify the amount of a physical quantity that is solely due to entanglement, a comparison to a separable evolution is required.
    In practice, this is often done by moving into a parameter-regime in which the state of the system becomes separable.
    However, such an approach leaves the precise contribution of entanglement ambiguous because changing physical parameters does often alter more than just entanglement.
    Thus, the precise role of entanglement in quantum thermodynamics, e.g., in cooling \cite{CPA13,BH14} and extracting work \cite{AF13} from of a system, is still an open area of research.

    In this contribution, the sharp dichotomy between quantum and classical correlations is explored by devising a general methodology for the classification of the dynamics of systems interacting with their environment, i.e., open quantum systems, regarding their entangling and disentangling properties.
    Specifically, a method to obtain the separable \cite{VP97,BN98,CV20} evolution is developed. 
    Comparing this evolution with the unrestricted, generally inseparable propagation in time allows one to determine the impact of entanglement over the duration of arbitrary noisy processes, even if the initial and final states do not carry any entanglement.
    This is a common situation in quantum computation, where an initially separable register of qubits is transformed by (noisy) entangling gates until final measurement projects onto the computational basis \cite{D00}.
    Our method extends the Monte Carlo wave function approach to yield a separable evolution.
    This is achieved by virtue of projections to the tangential space of separable states.
    Unlike many conventional treatments, we do not employ input-output relations but impose separability at every instant of solving the equations of motion.
    Thus, our method renders it possible to suppress entanglement while maintaining local superpositions and preserving classical correlations.
    Therefore, our approach allows for a genuine characterization of entanglement, without requiring a separable reference evolution, whose choice would be arbitrary.
    Based on our findings, we also derive novel types of nonlinear Lindblad master equations, consistently describing the separable dynamics, dubbed separability Lindblad equation. 
    This generalizes the two-qubit scenario in our accompanying work \cite{PAH24}.

    Nonlinear master equations are a rich and contemporary field of research by itself, originally used to describe the evolution of an effective single-particle density for an interacting many-body system \cite{L66,AM83}.
    Recently, nonlinear Lindblad master equations were employed to allow for arbitrary signs in the decay rates (e.g., for amplification) while ensuring complete positivity \cite{HJ23}.
    Another notable approach was the development of a correlation picture for open quantum systems \cite{AR20}, in which nonlinearities arise, due to initial system-bath correlations \cite{JS04,SS05}.

    The structure of the article is as follows. 
    In Sec. \ref{sec:NonSepDyn}, we recall the Lindblad master equation and the Monte Carlo wave function method.
    Section \ref{sec:SepMCWF} then establishes the separable Monte Carlo wave function approach, the key result of this paper.
    After applying the method in Sec. \ref{sec:Applications}, we derive the separability Lindblad equation in Sec. \ref{sec:SepMas}.
    This equation corresponds to a piece-wise deterministic process, which is equivalent to a stochastic equation.
    Finally, Sec. \ref{sec:Fin} contains a summary of the article as well as some concluding remarks.

\section{Preliminaries}
\label{sec:NonSepDyn}

    In this section, we revisit the conventional treatment of the time evolution of open quantum systems.
    Specifically, we briefly revisit the equation of motion, the notion of a separable process map, selected transformation properties, and the Monte Carlo wave function approach.
    Throughout this work, the system under study is considered to be of an arbitrary finite dimension and to consist of an arbitrary number of parties.

\subsection{Lindblad master equation and properties}

    Consider a time-dependent quantum state $\rho=\rho(t)$ subject to a quantum master equation in Lindblad form   
	\begin{equation}
    \begin{aligned}
        \label{eq:Lindblad}
        \frac{d\rho}{dt}
        &=i\left[\rho,H\right]
        +\sum_{a=1}^m\left(
            L^{a}\rho L^{a\dag}-\frac{1}{2}\left\{L^{a\dag} L^{a},\rho\right\}
        \right),
    \end{aligned}
	\end{equation}
    where $[\,\cdot\,,\,\cdot\,]$ and $\{\,\cdot\,,\,\cdot\,\}$ denote the commutator and anti-commutator, respectively, and $\hbar=1$.
    In Eq. \eqref{eq:Lindblad}, the Hamiltonian $H$ of the system yields a unitary contribution to the overall evolution of the system, and $L^{a}$ are Lindblad operators accounting for dissipative effects, which are labeled with the superscript $a$.
    The solution $\rho(t)=\Lambda_t(\rho(0))$ defines a linear map $\Lambda_t$, as discussed later.

    For a short propagation time $\tau$, the solution of Eq. \eqref{eq:Lindblad} leads to a (not necessarily separable) evolution
	\begin{equation}
        \label{eq:FormalSolution}
    \begin{aligned}
          \rho(t+\tau)
          =&\rho(t)+\tau\frac{d\rho(t)}{dt}+\mathcal O(\tau^2)\\
          =&\sum_{b=0}^m K^{b}\rho(t)K^{b\dag}+\mathcal O(\tau^2),
    \end{aligned}
	\end{equation}
	with the Kraus operators that read
	\begin{equation}
        \label{eq:KrausOps}
        K^{0}=\mathbbm{1} +\tau \left(
            \frac{1}{i}H-\frac{1}{2}\sum_{a=1}^m L^{a\dag} L^{a}
        \right)
        \quad\text{and}\quad
        K^{a}=\sqrt{\tau}L^{a},
	\end{equation}
    for $a\in\{1,\ldots,m\}$ and the $n$-partite identity $\mathbbm 1$.
    Furthermore, recall that Eq. \eqref{eq:Lindblad} remains invariant under the following transformations \cite{BP02}:
	\begin{equation}
    \begin{aligned}
		\label{eq:Substitution}
        H\mapsto {}&
        H+\sum_{a=1}^m\frac{\lambda^{a\ast} L^{a}-\lambda^{a}L^{a\dag}}{2i}
        \\
        \text{and}\quad
        L^{a}\mapsto {}&
        \lambda^{a}\mathbbm 1+L^{a},
    \end{aligned}
	\end{equation}
    with complex numbers $\lambda^{a}$.
    This symmetry allows us to write the Kraus operators in Eq. \eqref{eq:KrausOps} in a unified manner, viz.
    \begin{equation}
		\label{eq:perturbationForm}
		  K^b\mapsto K^b=\mu^b\left(\mathbbm 1+\varepsilon F^b\right),
	\end{equation}
    for $b\in\{0,\ldots,m\}$.
    Therein, we introduced the quantities $\mu^0=1$, $F^0=G/\|G\|$, and $\varepsilon=\tau\|G\|$, with the non-Hermitian generator
    \begin{equation}
        G=\frac{1}{i}H -\frac{1}{2}\sum_{a=1}^m\left(|\lambda^{a}|^2\mathbbm 1+2\lambda^{a\ast}L^{a}+L^{a\dag} L^{a}\right)
    \end{equation}
    and the operator norm $\|\,\cdot\,\|$,
    together with the quantities
    $\mu^a=\sqrt{\tau}\lambda^a$, $F^a=L^a/\|L^{a}\|$, and $\lambda^a=\|L^a\|/\varepsilon$ for $a\in\{1,\ldots,m\}$.
    In other words, for small $\varepsilon>0$, the Kraus operators in Eq. \eqref{eq:perturbationForm} can be expressed as a perturbation of the identity.
    This specific form of the Kraus operators turns out to be useful in our derivation in Sec. \ref{sec:SepMCWF}.
 
\subsection{Monte Carlo wave-function approach}

    Equivalent to the master equation treatment is the Monte Carlo wave function technique \cite{MCD93}.
    The method propagates an ensemble of wave functions, rather than the full density matrix $\rho$.
    With the Monte Carlo technique, the dynamics of each wave function is computed by randomly, propagating the ensemble of states via a non-Hermitian Hamiltonian and the Lindblad operators $L^a$ (quantum jumps).
    In our notation, this corresponds utilizing Kraus operators as expressed in Eq. \eqref{eq:perturbationForm}.

    Assuming one input state $\rho(t)=|\psi(t)\rangle\langle\psi(t)|$, the state after time $\tau$ is obtained by averaging over the possible outcomes
	\begin{equation}
		\label{eq:MCWFpreStep}
		|\phi^{b}\rangle=K^{b}|\psi(t)\rangle,
	\end{equation}
    for $b\in\{0,\dots,m\}$, with respect to the probability distribution
    \begin{equation}
    \label{eq:Q}
        p^b=\frac{\langle\phi^{b}|\phi^b\rangle}{Q},
        \quad\text{where}\quad
        Q=\sum_{b=0}^m \langle\phi^{b}|\phi^b\rangle.
    \end{equation}
    Note that, in most scenarios, the evolution is dominated by the process $K^0$ with rarely occurring quantum jumps, i.e., $p^0\gg p^a$ for $a\neq 0$.
    The formal solution \eqref{eq:FormalSolution}, in first order of $\tau$, is then obtained by averaging and normalizing the states in Eq. \eqref{eq:MCWFpreStep},
    \begin{equation}
        \label{eq:EvolvedRho}
        \rho(t+\tau)=
         \sum_{b=0}^m p^b \frac{|\phi^{b}\rangle\langle\phi^{b}|}{\langle\phi^{b}|\phi^b\rangle}
         =
        \frac{1}{Q}\sum_{b=0}^m K^b\rho(t)K^{b\dag}.
    \end{equation}
    Thus, the algorithm from Ref. \cite{MCD93} operates by randomly implementing a time step $t\mapsto t+\tau$ of the form $|\psi(t+\tau)\rangle=K^{b}|\psi(t)\rangle/\sqrt Q$, where $b$ is distributed according to the probability mass function in Eq. \eqref{eq:Q}.
    For a sufficiently large sample size, the average over the individual $|\psi(t+\tau)\rangle\langle\psi(t+\tau)|$ yields the desired output state $\rho(t+\tau)$ in Eq. \eqref{eq:EvolvedRho}.
	
\section{Separable Monte Carlo wave function}
\label{sec:SepMCWF}

    The techniques presented thus far make no distinction between entangling and non-entangling processes.
    To gauge the impact of entanglement during an open-system evolution, a method is required that yields non-entangling dynamics (of the same system) for comparison.
    The derivation of such a method is carried out in this section.

    To achieve the aforementioned goal, we extend the Monte Carlo wave function approach by restricting the evolution to the set of separable states of $n$ subsystems.
    A generic separable state can be written as \cite{W89}
	\begin{equation}
		\label{eq:sepStates}
		\rho(t)=\frac{\sum\limits_x|\psi^{x}(t)\rangle\langle\psi^{x}(t)|}{\sum\limits_x\langle\psi^{x}(t)|\psi^{x}(t)\rangle},
	\end{equation}
	with not necessarily normalized, $n$-fold tensor-product states
	\begin{equation}
		\label{eq:TPvecs}
		|\psi^x(t)\rangle
        =|\psi_1^x\rangle\otimes\cdots\otimes|\psi_n^x\rangle
        =|\psi_1^x\ldots\psi_n^x\rangle.
	\end{equation}

    The general outline of our derivation is as follows: We describe the tangent space of the product states that form the separable ensemble state, Sec. \ref{subsec:TangentSpace}.
    Then, in Sec. \ref{subsec:TangentialMaps}, the action of Kraus operators is restricted to that tangent space, which keeps the propagated state within the separable set.
    The combination of these techniques with the Monte Carlo technique yields the sought-after non-entangling propagation in time, Sec. \ref{subsec:sepMCWFalgorithm}.

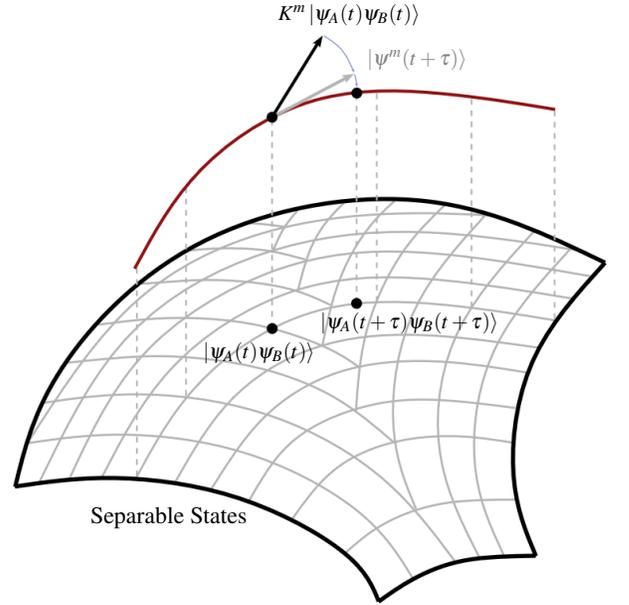
\begin{figure}
    \begin{tikzpicture}
        \node at (0,0) {\includegraphics[width=.5\textwidth]{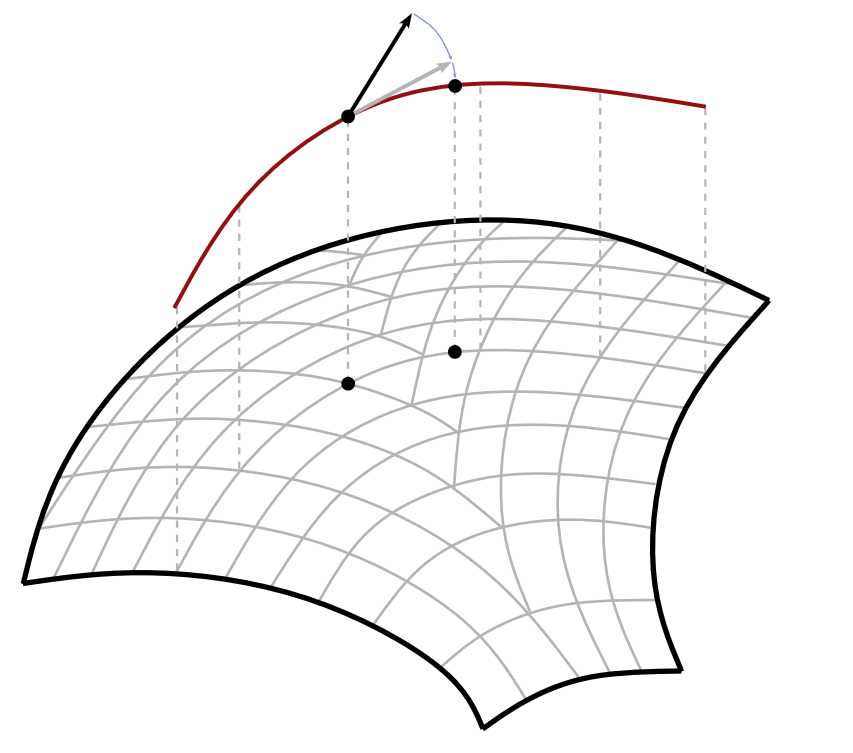}};
        \node at (-1,-0.4) {\footnotesize{$\ket{\psi_A(t)\psi_B(t)}$}};
        \node at (1,0) {\footnotesize{$\ket{\psi_A(t+\tau)\psi_B(t+\tau)}$}};
        \node at (1.1,3.5) {\footnotesize{\textcolor{gray}{$\ket{\psi^m(t+\tau)}$}}};
        \node at (0.2,4.1) {\footnotesize{$K^m\ket{\psi_A(t)\psi_B(t)}$}};
        \node at (-2.2,-2.6) {Separable States};
    \end{tikzpicture}
    \caption{%
        A Kraus operator $K^m$ takes the product state $\ket{\psi_A(t)\psi_B(t)}$ outside the manifold of separable states, cf. curve that is lifted from the manifold for improved visibility.
        A projection onto the tangent space, resulting in $\ket{\psi^\prime}=\ket{\psi^m(t+\tau)}$, of the manifold allows one to determine the updated product state $\ket{\psi_A(t+\tau)\psi_B(t+\tau)}$.
    }\label{fig:tangent}
\end{figure}

\subsection{Tangent space of product states}
\label{subsec:TangentSpace}

    Firstly, we specify the tangent space of pure product states.
    Later, we apply this to the full ensemble of states that defines the density operator, allowing us to preserve separability of mixed states within each time step (Fig. \ref{fig:tangent}).

    For the sake of simplicity, we assume that all subsystems are described by a $d$-dimensional Hilbert space, $\mathbb C^d$, each of which is described by the orthonormal basis $\{|0\rangle,\ldots,|d-1\rangle\}$.
    Expanding each factor in a product state $|\psi\rangle$ (we omit the state's label $x$ for ease of notation), $|\psi_k\rangle=\sum_{j=0}^{d-1}\psi_k^j|j\rangle$,  we can introduce tangent vectors as
	\begin{equation}
        \label{eq:tangentialTP}
		\frac{\partial|\psi\rangle}{\partial \psi^{j}_k}
		=|\psi_1\ldots\psi_{k-1}\rangle\otimes|j\rangle\otimes|\psi_{k+1}\ldots\psi_n\rangle,
	\end{equation}
    resulting in the $j$th basis element for the $k$th party, where $1\leq k\leq n$ and $0\leq j\leq d-1$.
    Consequently, an element of the tangent space at the product vector $|\psi\rangle$ can be identified with
    \begin{equation}
    \begin{aligned}
        \label{eq:TangentVec}
        {}&|\psi(t+\varepsilon)\rangle
        =|\psi\rangle
		+\sum_{k=1}^n \sum_{j=0}^{d-1}\varepsilon_k^{j}\frac{\partial|\psi\rangle}{\partial \psi^{j}_k}
        \\
        ={}&
        |\psi\rangle
        +|\varepsilon_1\psi_2\ldots\psi_n\rangle
        +\cdots
        +|\psi_1\ldots\psi_{n-1}\varepsilon_n\rangle,
    \end{aligned}
	\end{equation}
    with expansion coefficients $\varepsilon^j_k$, obeying $|\varepsilon^j_k|<\varepsilon$, and local vectors $|\varepsilon_k\rangle=\sum_{j=0}^{d-1}\varepsilon_k^j|j\rangle$ for $k\in\{1,\ldots,n\}$.
    These tangential elements can be related to product states as follows:
	\begin{equation}
    \begin{aligned}
		\label{eq:tangentialTPAprox}
        {}&|\psi(t+\varepsilon)\rangle
        \approx\bigotimes_{k=1}^n\left(|\psi_k\rangle+|\varepsilon_k\rangle\right)
		+\mathcal O(\varepsilon^2)
        \\
        ={}&|\psi_1\ldots\psi_N\rangle
        +\sum_{k=1}^n|\psi_1\ldots\psi_{k-1}\varepsilon_k\psi_{k+1}\ldots\psi_n\rangle
        +\mathcal O(\varepsilon^2),
    \end{aligned}
	\end{equation}
    which is accurate in the limit of small $\varepsilon$.
    The relation between the tangent vector \eqref{eq:TangentVec} and the product state \eqref{eq:tangentialTPAprox} is illustrated in Fig. \ref{fig:tangent} as well.

	In Eq. \eqref{eq:tangentialTPAprox}, the decomposition $|\varepsilon_k\rangle=|\varepsilon_{k}^{(\parallel)}\rangle+|\varepsilon^{(\perp)}_k\rangle$, with $|\varepsilon_{k}^{(\parallel)}\rangle\parallel|\psi_k\rangle$ and $|\varepsilon_{k}^{(\perp)}\rangle\perp|\psi_k\rangle$, may be used.
	While the parallel component $|\varepsilon^{(\parallel)}_k\rangle$ only contributes along the already present direction $|\psi_k\rangle$, $|\varepsilon_{k}^{(\perp)}\rangle$ truly alters the component.
	In addition, it is worth pointing out that, in Eq. \eqref{eq:tangentialTPAprox}, only one factor in the $k$th term, $|\psi_1\ldots\psi_{k-1}\varepsilon_k\psi_{k+1}\ldots\psi_n\rangle$, can include a $|\varepsilon_k^{(\perp)}\rangle$ contribution.
	Terms with more than one of such perpendicular factors are not in the tangent space and are referred to as $n$-orthogonal perturbations \cite{SV13}.

\begin{widetext}

\subsection{Maps and tangential projections}
\label{subsec:TangentialMaps}

	Next, we restrict the action of the maps under consideration, Eq. \eqref{eq:perturbationForm}, such that the output lies in the tangent space. 
    To this end, note that the action of any $K^b$ can be recast as follows
	\begin{equation}
		\label{eq:actionOfTildeKOnTP}
	\begin{aligned}
		K^b|\psi\rangle
		&=
		\mu^b\left( \mathbbm 1+\varepsilon
		\bigotimes_{k=1}^n\left(
			\frac{|\psi_k\rangle\langle\psi_k|}{\langle\psi_k|\psi_k\rangle}
			+\left[
			\mathbbm 1_k-\frac{|\psi_k\rangle\langle\psi_k|}{\langle\psi_k|\psi_k\rangle}
			\right]
		\right)
		F^b\right)
		|\psi_1\ldots\psi_n\rangle
		\\
		&= \mu^b\left(
			1+\varepsilon\frac{\langle\psi_1\ldots\psi_n|F^b|\psi_1\ldots\psi_n\rangle}{\langle\psi_1\ldots\psi_n|\psi_1\ldots\psi_n\rangle}
		\right)|\psi_1\ldots\psi_n\rangle\\
		&+\mu^b\varepsilon\sum_{k=1}^n
		|\psi_1\ldots\psi_{k-1}\rangle
		\otimes
			\left[
				\mathbbm 1_k
				-\frac{|\psi_k\rangle\langle\psi_k|}{\langle\psi_k|\psi_k\rangle}
			\right]\left(F^b\right)_k|\psi_k\rangle
		\otimes|\psi_{k+1}\ldots\psi_n\rangle
		+(\text{$n$-orthogonal perturbations}),
	\end{aligned}
	\end{equation}
	where we used the subsystem's identity $\mathbbm 1_k$ and so-called partially reduced operators,
	\begin{equation}
        \label{eq:RedOp}
		\left(F^b\right)_k=\frac{
			\left(
				\langle\psi_1\ldots\psi_{k-1}|
				\otimes\mathbbm 1_k\otimes
				\langle\psi_{k+1}\ldots\psi_n|
			\right)
			F^b
			\left(
				|\psi_1\ldots\psi_{k-1}\rangle
				\otimes\mathbbm 1_k\otimes
				|\psi_{k+1}\ldots\psi_n\rangle
			\right)
		}{
			\langle\psi_1\ldots\psi_{k-1}|\psi_1\ldots\psi_{k-1}\rangle
			\langle\psi_{k+1}\ldots\psi_n|\psi_{k+1}\ldots\psi_n\rangle
		}.
	\end{equation}
    From this definition follows the convenient identity
	\begin{equation}
        \label{eq:MeanFTP}
		\frac{\langle\psi_k|\left(F^b\right)_k|\psi_k\rangle}{\langle\psi_k|\psi_k\rangle}
		=\frac{\langle\psi|F^b|\psi\rangle}{\langle\psi|\psi\rangle}
		 =\langle F^b\rangle.
	\end{equation}
    Leaving out the $n$-orthogonal perturbations in Eq. \eqref{eq:actionOfTildeKOnTP} because they do not lie in the tangent space, we obtain the tangent vector $|\phi^b(t+\varepsilon)\rangle$.
    Up to first order in $\varepsilon$, the latter is equivalent to the product state
	\begin{equation}
		\label{eq:TPoutputGenericK}
	\begin{aligned}
        |\phi^b(t+\varepsilon)\rangle
        &\approx
        \mu^b \left(
			1+\varepsilon\langle F^b\rangle
		\right)
        \bigotimes_{k=1}^n\left(
			|\psi_k\rangle
			+\varepsilon
			\frac{
				\left[
				\mathbbm 1_k
				-\frac{|\psi_k\rangle\langle\psi_k|}{\langle\psi_k|\psi_k\rangle}
				\right]\left(F^b\right)_k|\psi_k\rangle
			}{
				1+\varepsilon\langle F^b\rangle
			}
		\right)
		\\
		&=
		\mu^b\frac{1}{\left(1+\varepsilon\langle F^b\rangle\right)^{n-1}}
		\bigotimes_{k=1}^n\left(\left[
			\mathbbm 1_k+\varepsilon \left(F^b\right)_k
		\right]
		|\psi_k\rangle\right)
		=\frac{1}{\langle K^b\rangle^{n-1}}
		\left(
			\bigotimes_{k=1}^n
			\left(K^b\right)_k
		\right)
		|\psi\rangle.
	\end{aligned}
	\end{equation}
    By doing so, we restrict the action of the operator $K^b$ to map onto the set of separable states.
    In other words, the partially reduced Kraus operators describes a non-entangling evolution as sought.

\end{widetext}

\subsection{The algorithm}
\label{subsec:sepMCWFalgorithm}

	Putting all previous findings together, one can devise a separability-preserving form of the Monte Carlo wave function approach.
	That is, for a given input $|\psi(t)\rangle$, one selects the product state  $|\phi^{b}\rangle/\sqrt{Q}$ according to Eq. \eqref{eq:TPoutputGenericK} and the probability mass function in Eq. \eqref{eq:Q}.
    Like in the original Monte Carlo approach, an average of a large enough sample of states yields a mixture after one time step,
	\begin{equation}
    \label{eq:SepMap}
        \rho(t+\tau)=\frac{1}{Q}\sum_{b=0}^m\frac{\left(\bigotimes_{k=1}^n \left(K^b\right)_{k}\right)\rho(t)\left(\bigotimes_{k=1}^n \left(K^b\right)_{k}\right)^\dag
		}{|\langle K^{b}\rangle|^{2(n-1)}}.
	\end{equation}

    It is insightful to compare Eq. \eqref{eq:SepMap} to one specific family of maps $\Lambda_t$, so-called separable maps \cite{VP97,BN98,CV20}.
    The dynamical map $\Lambda_t$ is said to be separable if it gives rise to an operator-sum representation of the form
    \begin{equation}
        \label{eq:DefSep}
        \Lambda_t(\rho)=\sum_{b=0}^m \big(K_{1}^b\otimes\cdots\otimes K_{n}^b\big) \rho \big(K_{1}^b\otimes\cdots\otimes K_{n}^b\big)^\dag,
    \end{equation}
    in which the time-dependent Kraus operators $K^b=K_{1}^b\otimes\cdots\otimes K_{n}^b$ are factorized.
    Notice that even though Eq. \eqref{eq:SepMap} has Kraus operators in product form, it corresponds to a nonlinear map, because the reduced Kraus operators $(K^b)_k$ depend explicitly on the state of the system, see Eq. \eqref{eq:RedOp}.
    Thus, it additionally guarantees the state $\rho(t+\tau)$ in Eq. \eqref{eq:SepMap} is separable, ensuring that our method describes the most general type of non-entangling dynamics.

    In contrast to the conventional Monte Carlo approach, here we are missing the unitary freedom in the choice of the Kraus operators. However, the separable dynamics remains invariant under the transformation
    \begin{equation}
        \frac{1}{\langle K^{b}\rangle^{n-1}}\bigotimes_{k=1}^n \big(K^b\big)_{k}\mapsto \sum_{c=0}^m\frac{U_{bc}}{\langle K^{c}\rangle^{n-1}}\bigotimes_{k=1}^n \big(K^c\big)_{k},
    \end{equation}
     with $U_{bc}$ being the components of a unitary $m\times m$ matrix.

\section{Applications}
\label{sec:Applications}

    The previously devised algorithm can be implemented straightforwardly alongside other solvers for the Lindblad equation.
    This allows us, for example, to study the entangling power of open-system processes when comparing our separability-preserving Monte Carlo approach with the common Monte Carlo wave function that can tap the resource of entanglement.
    In this section, a detailed study of essential examples is presented.

\begin{figure*}
    \includegraphics[width=\textwidth]{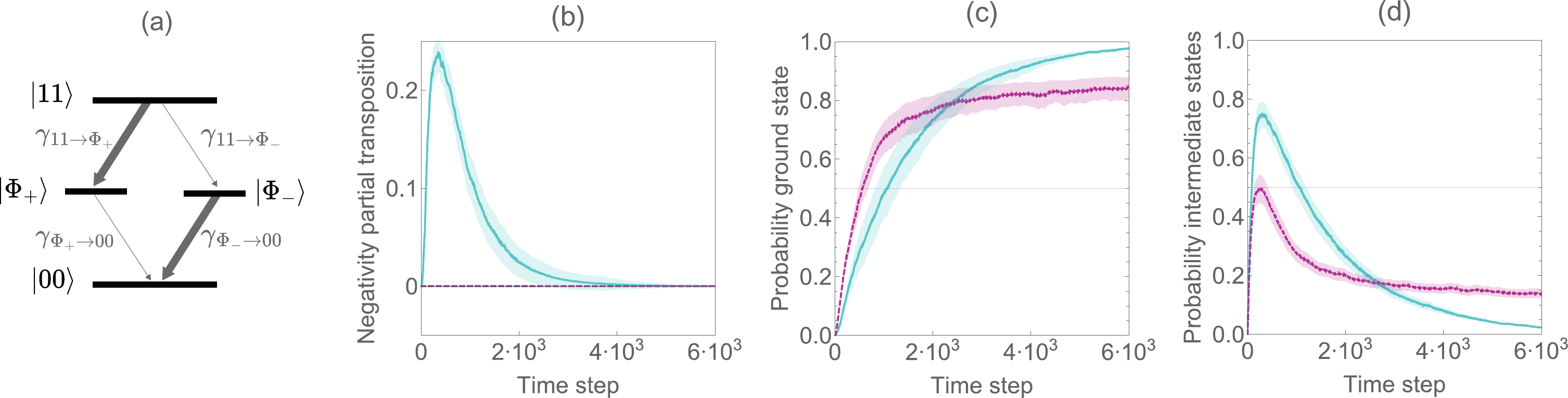}
    \caption{%
        \textbf{(a)} Scheme of the decay process given in Eq. \eqref{eq:exampleGenerators}.
        Comparison between the separable (dotted, purple line) and conventional (solid, turquoise line) Monte Carlo wave function methods:
		\textbf{(b)}
		Entanglement, characterized by the negativity of the partially transposed state.
		\textbf{(c)}
		Population of the ground state $\ket{00}$.
		\textbf{(d)}
		Probability to occupy one of the intermediate levels $|\Phi_+\rangle$ and $|\Phi_-\rangle$.
        One standard deviation uncertainty in lighter-colored bands.
        Note that the evolution of the probability to occupy the excited state can be computed by subtracting to $1$ the sum of probabilities displayed in (c) and (d).
		Parameters for the numerical simulation:
		$\gamma_{11\to\Phi_+}=9=\gamma_{\Phi_-\to00}, \gamma_{11\to\Phi_-}=1=\gamma_{\Phi_+\to00}$, step size 
		$\varepsilon=0.2$,
		and a sample size of $600$ \cite{Zenodo}. 
	}\label{fig:exampleLoss}
\end{figure*}

\subsection{Decay through meta-stable Bell states}
\label{subsec:ExampleCorrelatedLoss}

    We briefly discussed one example in our work \cite{PAH24}, which is studied here in more detail.
	In this example, which can be implemented in different physical platforms, \cite{VWC09,LGR13,PRD15}, we have a purely dissipative evolution, $H=0$, in which the initially doubly excited two-qubit state $|11\rangle$ decays into the ground state $|00\rangle$.
	The process is mediated by decay channels with metastable levels in the form of Bell states, $|\Phi_\pm\rangle=\tfrac{1}{\sqrt{2}}(|01\rangle\pm|10\rangle)$.
	Specifically, we have the following Lindblad operators,
	\begin{equation}
		\label{eq:exampleGenerators}
	\begin{aligned}
		&L^{1}=\sqrt{\gamma_{11\to\Phi_+}}|\Phi_+\rangle\langle 11|,\\
        &L^{2}=\sqrt{\gamma_{\Phi_+\to00}}|00\rangle\langle\Phi_+|,\\
        &L^{3}=\sqrt{\gamma_{11\to\Phi_-}}|\Phi_-\rangle\langle 11|,\\
        &L^{4}=\sqrt{\gamma_{\Phi_-\to00}}|00\rangle\langle\Phi_-|,
	\end{aligned}
	\end{equation}
	with the corresponding transition rates that obey $\gamma_{11\to\Phi_+}>\gamma_{11\to\Phi_-}$, such that most states decay into $|\Phi_+\rangle$, and $\gamma_{\Phi_+\to00}<\gamma_{\Phi_-\to00}$, resulting in a faster depletion of $|\Phi_-\rangle$;
	see Fig. \ref{fig:exampleLoss}(a).

	To evaluate Eq. \eqref{eq:SepMap}, first, the partially reduced operators from Eq. \eqref{eq:RedOp} are computed. For instance, one finds
	\begin{equation}
        \label{eq:PartReducedDecayBell}
        \begin{aligned}
		\big(L^1\big)_2&=\sqrt{\gamma_{11\to\Phi_+}}\frac{\left(\langle \psi_1|\otimes\mathbbm 1_2\right)\left(|\Phi_{+}\rangle\langle 11|\right)\left(|\psi_1\rangle\otimes\mathbbm 1_2\right)
		}{\langle \psi_1|\psi_1\rangle}\\
		&=\sqrt{\gamma_{11\to\Phi_+}}
		\frac{
  \psi_1^{1}}{\sqrt{2}}
		\frac{
			\big(\psi_1^1\big)^\ast|0\rangle
			+\big(\psi_1^0\big)^\ast|1\rangle
		}{
			|\psi_1^{0}|^2
			+|\psi_1^{1}|^2
		}
		\langle 1|,
        \end{aligned}
	\end{equation}
	where $|\psi_1\rangle=\psi_1^{0}|0\rangle+\psi_1^{(1)}|1\rangle$.
	One obtains similar for all partially reduced Lindblad operators.
    Because of the interaction between the subsystems, the partially reduced operator scales with the amplitude $|\psi_1^{1}|$ of the level $|1\rangle$ of the first qubit.
	Hence, the nonunitary  interactions between the subsystems are still taken into account, yet without introducing entanglement.
	In contrast to previous approaches, the method here allows one to compute and compare, for example, the efficiency of separable and inseparable processes in open quantum systems.
    To quantify the entanglement between the qubits, we compute the negativity of the partially transposed state \cite{VW02}, that is, the magnitude of its smallest negative eigenvalue.

    A comparison between the results of the separable and conventional Monte Carlo wave function approach is given in Fig. \ref{fig:exampleLoss}.
    As shown in panel (b), entanglement is generated during the decay but is not present over the full separable process; compare with Ref. \cite{BFP03}.
    Also the rate of decay into the separable ground state is different for the entangling and non-entangling scenarios, as shown in panel (c).
    Specifically, the population of the metastable, entangled states $|\Phi_+\rangle$ and $|\Phi_-\rangle$ [cf. panel (d)] is the reason for this behavior since the overlap, thus maximal population, of separable states with Bell states is bounded by $1/2$.
    For the same reason, the depletion is less efficient as the effective transition rate in Eq. \eqref{eq:PartReducedDecayBell} scales with the overlap of the Bell state with the separable one.
    See Ref. \cite{YS22} for a quantitative assessment of the speed of separable and inseparable dynamics.

\begin{figure*}
    \includegraphics[width=\textwidth]{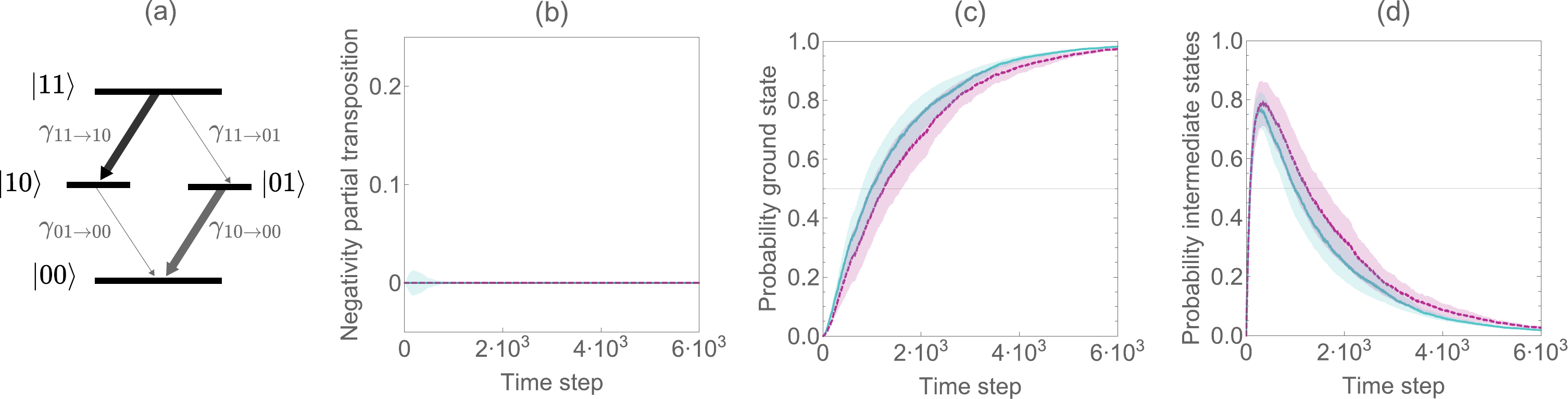}
    \caption{%
        Comparison between the separable (dashed, purple line) and conventional (solid, turquoise line) Monte Carlo wave function method for the decay process given in Eq. \eqref{eq:exampleGeneratorsSep}, with one standard deviation uncertainty (lighter colored bands).
        Contrasting the process in Fig. \ref{fig:exampleLoss}, the here described decay mechanism does not lead to entanglement, panel (b).
        Also, decay into the ground state, panel (c), and the occupation of the intermediate states, panel (d), show no significant deviation from another.
        This confirms the expected consistency of the separable Monte Carlo technique when no entanglement is involved in the process.
        Selected parameters for the numerical simulation are the same as in Fig. \ref{fig:exampleLoss}.}
    \label{fig:exampleLossSep}
\end{figure*}

\subsection{Decay through product states}
\label{subsec:ExampleCorrelatedLossSep}

    Contrasting the entangled intermediate levels of the previous example, we now analyze the case when both intermediate levels are separable states as well, namely $\ket{10}$ and $\ket{01}$.
    To this end, we define the Lindblad operators as
    \begin{equation}
		\label{eq:exampleGeneratorsSep}
	\begin{aligned}
		&L^{1}=\sqrt{\gamma_{11\to10}}|10\rangle\langle 11|,\\
        &L^{2}=\sqrt{\gamma_{10\to00}}|00\rangle\langle 10|,\\
        &L^{3}=\sqrt{\gamma_{11\to01}}|01\rangle\langle 11|,\\
        &L^{4}=\sqrt{\gamma_{01\to00}}|00\rangle\langle01|.
	\end{aligned}
	\end{equation}
    Contrasting the earlier case in Fig. \ref{fig:exampleLoss}, the complementary results from the scenario studied here are shown in Fig. \ref{fig:exampleLossSep}.
    As there is no entanglement present in the process-defining operators, one expects and the simulation confirms that both the separable and conventional Monte Carlo wave function approach give compatible trajectories within the margin of error from the inherent random fluctuations of the methods.
    Note that, in order to arrive at this result, it is not necessary for the Lindblad operators in Eq. \eqref{eq:exampleGeneratorsSep} to be given in product form.
    It is well known that applying a unitary transformation $L^a\mapsto \sum_{b=1}^m U_{ab}L^b$ yields the same physics; i.e., Eq. \eqref{eq:Lindblad} is invariant under such transformations.
    Despite the restricted dynamics in Eq. \eqref{eq:SepMap} being nonlinear, for a separable map, we find that the trajectories are compatible independent of the representation.
    For instance, set $\gamma=\gamma_{11\to10}=\gamma_{11\to01}$ and choose
    \begin{equation}
        \begin{split}
            L^1&\mapsto \frac{1}{\sqrt{2}}\big(L^1+L^3\big)=\sqrt{\gamma}\ket{\Phi^+}\bra{11}
            \\\text{and}\quad
            L^3&\mapsto \frac{1}{\sqrt{2}}\big(L^3-L^1\big)=\sqrt{\gamma}\ket{\Phi^-}\bra{11}\\
        \end{split}
    \end{equation}
    as new Lindblad operators, while leaving $L^2$ and $L^4$ unchanged.
    This makes it much harder to identify that the decay process is indeed separable.
    Still, applying the separable Monte Carlo wave function method yields again compatible trajectories with the unrestricted approach.

\begin{figure*}
    \includegraphics[width=\textwidth]{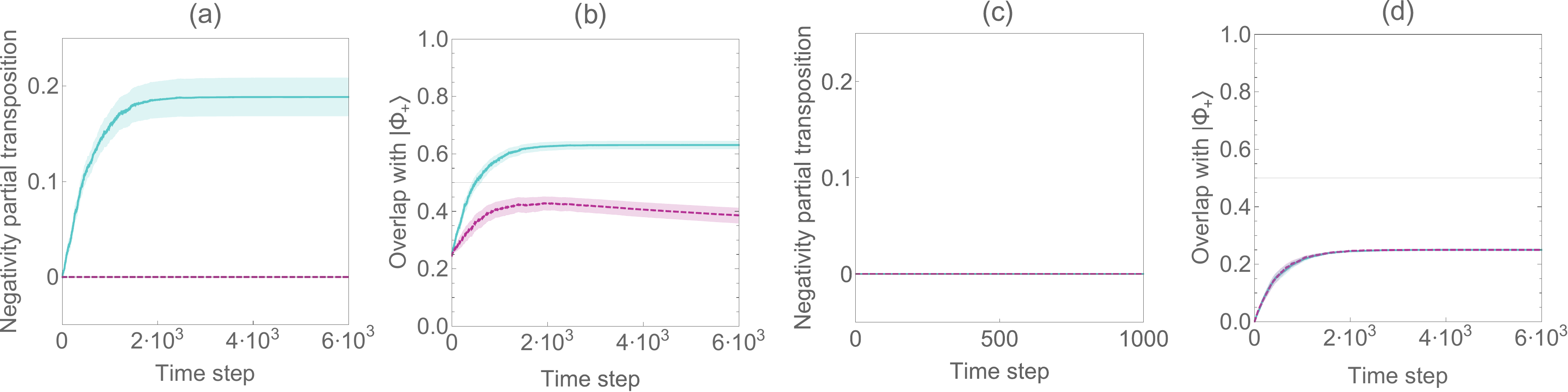}
    \caption{
        Comparison between the separable (dashed, purple line) and conventional (solid, turquoise line) Monte Carlo wave function method for the process in Eq. \eqref{eq:exampleCNOTGenerator} with one standard deviation uncertainty (lighter colored bands):
        \textbf{(a)}
        For the initial state $(|0\rangle+|1\rangle)/\sqrt2\otimes\ket{0}$, entanglement is quantified via negativity under partial transposition.
        \textbf{(b)}
        For the same initial state, overlap between $\rho(t)$ and the Bell state $\ket{\Phi_+}$.
        The trajectories do not converge due to the entanglement present of state for $t\to\infty$ for the unrestricted evolution.
        Panels \textbf{(c)} and \textbf{(d)} depict the results for the initial state $\ket{10}$.
        The expectation that both the separable and inseparable technique ought to lead to the same classically correlated outcome for this input is confirmed.
        The parameters for the numerical simulations are
        $\varepsilon=0.2$ and a sample size of $400$ \cite{Zenodo}.
    }\label{fig:exampleCNOT}
\end{figure*}

\subsection{Correlated bit-flip errors}
\label{subsec:ExampleCNOT}

    As a third example, we consider a model for bit-flip error that is induced by crosstalk to another bit, being described as a controlled NOT operation.
    We consider a purely dissipative process, $H=0$, the initial state of the bipartite system is $\frac{1}{\sqrt{2}}(|0\rangle+|1\rangle)\otimes\ket{0}$,
    and the open-system dynamics is mediated by the controlled NOT operator
    \begin{equation}
        \label{eq:exampleCNOTGenerator}
        L=\begin{pmatrix}
            1 & 0 & 0 & 0\\
            0 & 1 & 0 & 0\\
            0 & 0 & 0 & 1\\
            0 & 0 & 1 & 0
        \end{pmatrix}.
    \end{equation}

    A comparison between the Monte Carlo simulation for the separable and inseparable evolution is shown in Fig. \ref{fig:exampleCNOT}.
    Panel (a) illustrates the entanglement buildup during the unrestricted evolution, while the state, indeed, remains separable during the restricted evolution.
    In panel (b), we show the overlap between $\rho(t)$ and the Bell state $\ket{\Psi_+}=\tfrac{1}{\sqrt{2}}(|00\rangle+|11\rangle)$ that one expects when applying the controlled NOT to the input state.
    Again, this overlap is bounded by $1/2$ for the separable evolution and may exceeds this value for the unrestricted approach.
    Since the steady state of the system, being
    \begin{equation}
        \rho_\mathrm{s}=\frac{1}{2}\rho(0)+\frac{1}{2}L\rho(0)L^\dagger,
    \end{equation}
    is entangled for the initial state $\frac{1}{\sqrt{2}}(\ket{0}+\ket{1})\otimes\ket{0}$, the restricted evolution does not converge to $\rho_\mathrm{s}$, which is evident in panel (b) as the dashed curve does not converge to the same value as the solid one.

    Finally, we consider the initial state $|10\rangle$ with the expectation that the unrestricted evolution remains separable for all times.
    The result is depicted in panels (c) and (d) of Fig. \ref{fig:exampleCNOT}. 
    Indeed, for this initial state, the separable and unrestricted evolution coincide.
    While this is surprising since the controlled NOT is not a local operator, as mentioned in Sec. \ref{subsec:ExampleCorrelatedLossSep}, the Lindblad operator does not need to be in product form to generate separable dynamics.
    This becomes clear in the present scenario, cf. Eq. \eqref{eq:exampleCNOTGenerator}, where the trajectories coincide due to the choice of initial state.
    
    We observe that the separable Monte Carlo wave function approach does not only accurately capture the evolution of a separable map [cf. \eqref{eq:DefSep}], but coincides with the unrestricted (physical) evolution for any non-entangling process. 

\section{Equations of motion}
\label{sec:SepMas}

    In general, a stochastic process can be cast into equivalent equations of motions that are then solved by the algorithm;
    see Ref. \cite{MCD93}.
    Here, we derive the underlying Lindblad-type and stochastic differential equations to our separable Monte Carlo approach.
    In addition, further examples and properties are discussed in this section.


\subsection{Derivation of the separability Lindblad equation}

    Recall that, according to the separable Monte Carlo method, the propagation of an initial product state $\rho(t)=\bigotimes_{k=1}^n\ket{\psi_k(t)}\bra{\psi_k(t)}$ is governed by the separable (nonlinear) map in Eq. \eqref{eq:SepMap}.
    Here, the reduced Kraus operators can be written as
    $\big(K^{0}\big)_k=\mathbbm{1}_k-i \tau \big(\hat{H}\big)_k,$ and $\big(K^{a}\big)_k=\sqrt{\tau }\big(L^{a}\big)_k$, with $\hat{H}=H-\tfrac{i}{2}\sum_{a=1}^m L^{a\dag} L^{a}$ being an effective non-Hermitian Hamiltonian.
 
\begin{widetext}
    Inserting this form of the Kraus operators in Eq. \eqref{eq:SepMap} and keeping only contributions up to first order in $\tau$ yields
    \begin{equation}
        \begin{split}
            Q\rho(t+\tau)&\approx \rho-i\tau(n-1)\big(\braket{\hat{H}}-\braket{\hat{H}}^*\big)\rho-i\tau\left(\sum_{k=1}^n \mathbbm{1}^{\otimes (k-1)}\otimes \big(\hat{H}\big)_k\otimes \mathbbm{1}^{\otimes (n-k)}\right)\rho\\
            &+i\tau\rho\left(\sum_{k=1}^n \mathbbm{1}^{\otimes (k-1)}\otimes \big(\hat{H}\big)_k^\dag\otimes \mathbbm{1}^{\otimes (n-k)}\right)+\tau \sum_{a=1}^m \frac{1}{|\braket{L^{a}}|^{2(n-1)}}\bigotimes_{k=1}^n \big(L^{a}\big)_k\rho \bigotimes_{k=1}^n \big(L^{a}\big)_k^\dag,\\
        \end{split}
    \end{equation}
    where from now on $\mathbbm 1$ denotes the identity on the individual subsystem and $Q$ was the normalizing constant in Eq. \eqref{eq:SepMap}.
    Making use of $Q\rho(t+\tau)\approx \rho+\tau \mathcal{L}_{\rho}(\rho)$ and the explicit form of $\hat{H}$, gives us the generator of separable dynamics $\mathcal{L}_\rho$,
    \begin{equation}
        \label{eq:SepGen}
        \begin{split}
        \mathcal{L}_\rho(\rho)&= i\sum_{k=1}^n\left[\rho,\mathbbm{1}^{\otimes (k-1)}\otimes (H)_k\otimes \mathbbm{1}^{\otimes (n-k)}\right]-\frac{1}{2}\sum_{a=1}^m\sum_{k=1}^n\Big\{\mathbbm{1}^{\otimes (k-1)}\otimes\Big(L^{a\dag} L^{a}\Big)_k\otimes \mathbbm{1}^{\otimes (n-k)},\rho\Big\}\\
        &+(n-1)\left\langle\sum_{a=1}^m L^{a\dag} L^{a}\right\rangle\rho+\sum_{a=1}^m \frac{1}{|\braket{L^{a}}|^{2(n-1)}}\bigotimes_{k=1}^n \big(L^a\big)_k\rho\bigotimes_{k=1}^n \big(L^a\big)_k^\dag.\\
        \end{split}
    \end{equation}
\end{widetext}

    Subtracting the initial state from the evolved state and noting that $\tfrac{{d}}{{d} t}\rho\approx\tfrac{Q\rho(t+\tau)-\rho(t)}{\tau}$ for small $\tau$, yields the equation of motion for the separable dynamics
    \begin{equation}
        \label{eq:SepMaster}
        \frac{{d} \rho}{{d} t}=\mathcal{L}_\rho(\rho).
    \end{equation}
    We refer to this equation of motion for the generator $\mathcal L_\rho$ in Eq. \eqref{eq:SepGen} as the separability Lindblad equation.
    Equation \eqref{eq:SepMaster} is nonlinear since both the partially reduced operators $(L^m)_k$ and $(H)_k$, for $k=A,B$, as well as the mean values depend explicitly on the state $\rho$.
    Thereby, it describes the evolution of a composite system restricted to a separable trajectory, which are ensured by the tangential projections discussed earlier and resulting in the nonlinear properties.
    The procedure removes any entanglement from the full evolution [subject to Eq. \eqref{eq:Lindblad}], while preserving all classical correlations and interactions \cite{PM24}.
    Comparing the solution of the separability Lindblad equation with those of the regular Lindblad equation, we are able to study the inseparable dynamics of compound quantum systems.

    Importantly, Eq. \eqref{eq:SepMaster} should be understood as a formal expression as it is not a deterministic equation.
    This is due to the fact that the generator $\mathcal{L}_\rho$ is defined for a pure product state $\rho$ only.
    In general, dissipation leads to $\rho$ evolving into a mixed state after an infinitesimal time step.
    The evolved state is then obtained via a mixture of pure-state trajectories, i.e., $\rho(t+\tau)=\sum_x p_x \rho^x(t+\tau)$, with pure product states $\rho^x(t+\tau)=\bigotimes_k\ket{\psi^x_k}\bra{\psi^x_k}$, allows one to propagate each of these states via 
    \begin{equation}                
        \rho^x(t+2\tau)=\rho^x(t+\tau)+\tau\mathcal{L}_{\rho^x(t+\tau)}\big(\rho^x(t+\tau)\big),
    \end{equation}
    being the essence of the separability Monte Carlo approach devised above.
    In this sense, the separability Lindblad equation \eqref{eq:SepMaster} describes a piece-wise deterministic process \cite{BP02,D93}, in which the generator $\mathcal{L}_\rho$ has to be updated after each step.

\subsection{Consistency and special cases}

    Given any dynamical map, the form of the separability Lindblad equation ensures an initially separable state to remain separable during its evolution.
    For such an approach to be physically meaningful, it is important that, for a separable map, the unrestricted dynamics [Eq. \eqref{eq:Lindblad}] and the restricted dynamics [Eq. \eqref{eq:SepMaster}] are identical.
    In Appendix \ref{app:Cons}, we show that this is indeed the case.
    Furthermore, for a single subsystem, the separability Lindblad equation in Eq. \eqref{eq:SepMaster} is equivalent to the conventional one, which can be directly seen when carefully evaluating the generator in Eq. \eqref{eq:SepGen} for $n=1$.

    Moreover, for a closed quantum system the separability Lindblad equation reduces to the separability Schr\"odinger equations introduced in Ref. \cite{SW17}.
    This can be concluded from Eq. \eqref{eq:SepMaster}, which for $L^a=0$, with $a=1,\dots,m$, becomes
    \begin{align}
        \nonumber
        i\frac{{d}\rho}{{d} t}
        ={}&
        i\sum_{k=1}^n\bigotimes_{k=1}^{j-1}\ket{\psi_k}\bra{\psi_k}{\otimes} \frac{{d}}{{d} t}\big(\ket{\psi_j}\bra{\psi_j}\big)
        {\otimes} \bigotimes_{k=j+1}^n\ket{\psi_k}\bra{\psi_k}
        \\
        ={}&\sum_{k=1}^n\left[\mathbbm{1}^{\otimes (k-1)}\otimes (H)_k\otimes \mathbbm{1}^{\otimes (n-k)},\rho\right].
    \end{align}
    Tracing over all subsystems but the $n$th one, leads to a set of coupled von Neumann-type equations $i\frac{{d}}{{d} t}\big(\ket{\psi_j}\bra{\psi_j}\big)=\big[(H)_{j},\ket{\psi_j}\bra{\psi_j}\big]$,
    where $(H)_j$ is the reduced Hamiltonian of the system obtained from Eq. \eqref{eq:RedOp}.
    When ignoring global phases, these von Neumann-type equations are equivalent to the separability Schr\"odinger equations from Ref. \cite{SW17}.

\subsection{Example: random exchange interaction}

    In terms of our equations of motion, let us consider a purely dissipative process in a bipartite system, i.e., $H=0$, with a single Lindblad operator $L=\sqrt{\gamma} V$, where $V\ket{\psi_1\psi_2}=\ket{\psi_2\psi_1}$ is the swap.
    The unrestricted evolution is governed by the master equation 
    \begin{equation}
        \label{eq:FullSwap}
        \frac{{d} \rho}{{d} t}=\gamma V\rho V -\gamma \rho,
    \end{equation}
    where we made use of $V=V^\dag$ and $V^2=\mathbbm{1}^{\otimes 2}$.
    The dynamical map for this case can be given analytically via exponentiation of the generator in Eq. \eqref{eq:FullSwap}, viz. 
    \begin{equation}
        \label{eq:FullSwapSol}
        \rho(t)={e}^{-\gamma t}\big(\cosh(\gamma t)\rho+\sinh(\gamma t)V\rho V\big).
    \end{equation}
    The evolution in Eq. \eqref{eq:FullSwapSol} corresponds to a random unitary map \cite{AS08}, in which with probability $\sinh(\gamma \tau){e}^{-\gamma \tau}$ the swap operation is applied.
    While the swap operation does not create entanglement, it is still non-separable; i.e., its Kraus operators cannot be brought into product form.
    While these maps cannot introduce entanglement on their own, they have the capacity to generate multipartite entanglement when coupled to subsystems in which entanglement is already present \cite{VH05}.

    In order to obtain the restricted dynamics, we seek a solution of the separability master equation \eqref{eq:SepMaster}.
    First, note that
    \begin{equation}
        \begin{split}
            (V)_1\ket{\psi_1}=\frac{\big(\mathbbm{1}\otimes\bra{\psi_2}\big)V\ket{\psi_1\psi_2}}{\braket{\psi_2|\psi_2}}&=\frac{\braket{\psi_2\vert\psi_1}}{\braket{\psi_2|\psi_2}} \ket{\psi_2},\\
            (V)_2\ket{\psi_2}=\frac{\big(\bra{\psi_1}\otimes\mathbbm{1}\big)V\ket{\psi_1\psi_2}}{\braket{\psi_1|\psi_1}}&=\frac{\braket{\psi_1\vert\psi_2}}{\braket{\psi_1|\psi_1}} \ket{\psi_1},\\
        \end{split}
    \end{equation}
    implies $(V)_1=\tfrac{\ket{\psi_2}\bra{\psi_2}}{\braket{\psi_2|\psi_2}}$ and $(V)_2=\frac{\ket{\psi_1}\bra{\psi_1}}{\braket{\psi_1|\psi_1}}$.
    It follows that the reduced operators for the present case are
    \begin{equation}
            \begin{split}
             (L)_1&=\sqrt{\gamma}\frac{\ket{\psi_2}\bra{\psi_2}}{\braket{\psi_2|\psi_2}},\quad (L^\dag L)_1=\gamma\mathbbm{1},\\
             (L)_2&=\sqrt{\gamma}\frac{\ket{\psi_1}\bra{\psi_1}}{\braket{\psi_1|\psi_1}},\quad (L^\dag L)_2=\gamma\mathbbm{1}.
            \end{split}
    \end{equation}
    Inserting these operators into Eq. \eqref{eq:SepMaster} leads us to 
    \begin{equation}
        \label{eq:RedSwap}
        \begin{split}
            \frac{{d} \rho}{{d} t}&=\frac{\gamma}{\braket{V}^2}(V)_1\otimes (V)_2 \rho (V)_1\otimes (V)_2-\gamma \rho,\\
        \end{split}
    \end{equation}
    where $\braket{V}=|\braket{\psi_1|\psi_2}|^2$.
    Equation \eqref{eq:RedSwap} has to be solved iteratively starting from the normalized state $\rho=\ket{\psi_1\psi_2}\bra{\psi_1\psi_2}$. 
    After the first time step, the state propagates to
    \begin{equation}
             \rho(\tau)= 
             (1-\gamma\tau)\ket{\psi_1\psi_2}\bra{\psi_1\psi_2}+ \gamma \tau\ket{\psi_2\psi_1}\bra{\psi_2\psi_1},
    \end{equation}
    for sufficiently small $\tau$.
    The procedure continues for states at later time steps $s\tau$, with $s>2$.
    In this particular example, the separability master equation can be solved analytically, viz.
    \begin{equation}
        \label{eq:SolSwap}
        \begin{split}
            \rho(s\tau)=&\sum_{a\text{ even}}^{s}{s\choose a}(1-\gamma \tau)^{s-a}(\gamma\tau)^a\ket{\psi_1\psi_2}\bra{\psi_1\psi_2}\\
            &+\sum_{a\text{ odd}}^{s}{s\choose a}(1-\gamma \tau)^{s-a}(\gamma\tau)^a\ket{\psi_2\psi_1}\bra{\psi_2\psi_1}.
        \end{split}
    \end{equation}
    The sums over even and odd indices in Eq. \eqref{eq:SolSwap}, for $\tau=t/s$, can be written as 
    \begin{equation}
    \begin{aligned}
        \sum_{a=0}^{s}{s\choose 2a}(1-\gamma \tau)^{s-2a}(\gamma \tau)^{2a}
        =&{}
        \frac{
            1{+}\Big(1 {-} 2\frac{\gamma t}{s}\Big)^s
        }{2},
        \\
        \sum_{a=0}^{s}{s\choose 2a+1}(1-\gamma \tau)^{s-(2a+1)}(\gamma \tau)^{2a+1}
        ={}&
        \frac{
            1{-}\Big(1 {-} 2\frac{\gamma t}{s}\Big)^s
        }{2},
    \end{aligned}
    \end{equation}
    respectively.
    Taking the limit $s\to\infty$ yields
    ${e}^{-\gamma t}\cosh(\gamma t)$ and ${e}^{-\gamma t}\sinh(\gamma t)$, respectively.
    This shows that the restricted dynamics in Eq. \eqref{eq:SolSwap} coincide with the unrestricted dynamics in Eq. \eqref{eq:FullSwapSol}.

    The above example highlights that the separability Lindblad equation \ref{eq:SepMaster} covers all aspects of non-entangling evolutions not just dynamical maps, whose Kraus operators are in product form, i.e., the separable maps defined in Eq. \eqref{eq:DefSep}.
    While the latter can be identified using the Choi matrix \cite{C75,J72} of the map, so far, there was no method to distinguish general non-entangling processes from ones which can create entanglement \cite{CV20}.
    The separability Lindblad equation provides a unique method to achieve this goal. 

\subsection{Stochastic separability equations}
\label{sec:Stoch}
    
    Mathematically speaking, the Monte Carlo wave function approach corresponds to a piece-wise deterministic process \cite{BP02}.
    The dynamics given by a piece-wise deterministic process can also be obtained from an equivalent stochastic differential equation.
    In this formulation, a single quantum trajectory $\ket{\psi(t)}$ evolves according to the non-Hermitian Hamiltonian
    $\hat{H}=H-\frac{i}{2}\sum_{a=1}^m L^{a\dag} L^{a}$ interrupted by random quantum jumps $\ket{\psi}\mapsto L^a\ket{\psi}$;
    see Eq. \eqref{eq:KrausOps}.
    The stochastic Schr\"{o}dinger equation reads \cite{BP02,WM93}
    \begin{equation}
        \label{eq:stoch}
        \begin{split}
            {d}\ket{\psi}=&\left(\frac{1}{i}\hat{H}+\frac{1}{2}\sum_{a=1}^m\braket{L^{a\dag}L^a}\right)\ket{\psi}{d} t\\
            &+\sum_{a=1}^m\left(\frac{L^a\ket{\psi}}{\braket{L^{a\dag}L^a}^{1/2}}-\ket{\psi}\right)dN_a,
        \end{split}
    \end{equation}
    where $dN_a(t)=1$ (and ${d} t=0$) at the instants of time where the measurement outcome $a$ is obtained due to nonselective measurements by the environment.
    Otherwise, we have ${d}N_a(t)=0$.
    Moreover, the Poisson increments satisfy
    \begin{equation}
        \label{eq:Pois}
        {d} N_a(t){d} N_b(t)=\delta_{ab}{d} N_b(t)
        \text{ and }
        \mathbbm{E}[{d} N_a(t)]=\braket{L^{a\dag}L^a}{d} t.
    \end{equation}
    Note that Eq. \eqref{eq:stoch} is nonlinear because of the mean values $\braket{L^{a\dag}L^a}=\bra{\psi}L^{a\dag}L^a\ket{\psi}$ depending on the form of $\ket{\psi}$.

    Equation \eqref{eq:stoch} can also be formulated on the level of density operators.
    The corresponding stochastic von Neumann equation reads
    \begin{equation}
        \label{eq:stoch2}
        \begin{split}
            {d}\sigma=&i\left(\sigma\hat{H}^\dag-\hat{H}\sigma\right){d} t+\sum_{a=1}^m\braket{L^{a\dag}L^a}\sigma{d} t\\
            &+\sum_a\left(\frac{L^a\sigma L^{a\dag}}{\braket{L^{a\dag}L^a}}-\sigma\right){d}N_a.
        \end{split}
    \end{equation}
    The solution of Eq. \eqref{eq:stoch2} can be obtained via iteration. 
    Given an initial condition $\sigma_0$, we obtain the state at a later time from ${d} \sigma=\sigma_{t+1}-\sigma_t$.
    It should be emphasized that the stochastic von Neumann equation \eqref{eq:stoch2} preserves the purity of the state $\sigma$.
    For instance, starting from a pure state $\sigma_0=\ket{\psi_0}\bra{\psi_0}$ at time $t=0$, the result of an open system evolution is obtained via an averaging over many trajectories, i.e., $\rho(t)=\mathbb{E}[\sigma_t]$.
    When averaging Eq. \eqref{eq:stoch2} over many trajectories, while making use of ${d}\rho=\mathbb{E}[{d}\sigma]$ and $\mathbb{E}[{d} N_a(t)]=\braket{L^{a\dag}L^a}{d} t$ [see Eq. \eqref{eq:Pois}], one recovers the Lindblad master equation \eqref{eq:Lindblad}.

    Following the main theme of this paper, we can introduce separability-preserving analogues of Eqs. \eqref{eq:stoch} and \eqref{eq:stoch2}.
    Treating the separable Monte Carlo method \eqref{eq:SepMap} as a piece-wise deterministic process, the equivalent stochastic (separability) Schr\"{o}dinger equation reads
    \begin{equation}
        \label{eq:stochSep}
        \begin{split}
            {d}\ket{\psi}=&\frac{1}{i}\sum_{k=1}^n\mathbbm{1}^{\otimes(k-1)}\otimes(\hat{H})_k\otimes\mathbbm{1}^{\otimes (n-k)}\ket{\psi}{d} t\\
            &-\left(\frac{n-1}{i}\braket{\hat{H}}+\frac{1}{2}\sum_{a=1}^m\braket{L^{a\dag}L^a}\big)\right)\ket{\psi}{d} t\\
            &+\sum_{a=1}^m\left(\frac{\bigotimes_{k=1}^n(L^a)_k\ket{\psi}}{\braket{L^a}^{n-1}\braket{L^{a\dag}L^a}^{1/2}}-\ket{\psi}\right){d}N_a,\\
        \end{split}
    \end{equation}
    where $\ket{\psi}=\bigotimes_{k=1}^n\ket{\psi_k}$ remains in product form throughout the evolution.
    For a local Hamiltonian [see Eq. \eqref{eq:LocOp}]
    \begin{equation}
        \hat{H}=\sum_{k=1}^{n}\mathbbm{1}^{\otimes(k-1)}\otimes \hat{H}_k\otimes \mathbbm{1}^{\otimes(n-k)},
    \end{equation}
    and factorizable jump operators $L^a=\bigotimes_{k=1}^{n} L^a_k$, Eq. \eqref{eq:stochSep} reduces to the stochastic Schr\"{o}dinger equation \eqref{eq:stoch} of a separable map [see Eq. \eqref{eq:RedSep}].

    We can formulate an equivalent von Neumann-type equation for the pure product state $\sigma_t=\ket{\psi_t}\bra{\psi_t}$.
    That is,
    \begin{equation}
        \label{eq:stochSep2}
        \begin{split}
            {d} \sigma=&\left(\frac{1}{i}\sum_{k=1}^n\left[\mathbbm{1}^{\otimes(k-1)}\otimes(H)_k\otimes\mathbbm{1}^{\otimes (n-k)},\sigma \right]\right.\\
            &-\frac{1}{2}\sum_{a=1}^m\sum_{k=1}^n\left\{\mathbbm{1}^{\otimes(k{-}1)}{\otimes}\big(L^{a\dag}L^a\big)_k{\otimes}\mathbbm{1}^{\otimes (n-k)},\sigma \right\}\\
            &\left.+n\sum_{a=1}^m\braket{L^{a\dag}L^a}\sigma \right){d} t\\
            &+\sum_{a=1}^m\left(\frac{\otimes_{k=1}^n(L^a)_k\sigma\otimes_{k=1}^n(L^a)_k^\dag}{|\braket{L^a}|^{2(n-1)}\braket{L^{a\dag}L^a}}-\sigma\right){d} N_a.
        \end{split}
    \end{equation}
    When averaging Eq. \eqref{eq:stochSep2} over many trajectories, while making use of ${d}\rho=\mathbb{E}[{d}\sigma]$ and $\mathbb{E}[{d} N_a(t)]=\braket{L^{a\dag}L^a}{d} t$ [see Eq. \eqref{eq:Pois}], one recovers the separability Lindblad equation \eqref{eq:SepMaster}.

\section{Conclusions}
\label{sec:Fin}

    We derived an approach to restrict the dynamics in open systems to non-entangled states, which allows us to determine the impact of entanglement in unrestricted open-system processes.
    This was achieved by forcing the Monte Carlo wave function approach to yield separable trajectories through projections onto the tangent space of the manifold of product wave functions.
    From sampling over a large number of trajectories, classical correlations arise without the buildup of entanglement.
    Contrasting this with the unrestricted evolution provides a necessary and sufficient criteria for multipartite entanglement in open quantum systems for arbitrary dimensions.

    Several examples showed the potential of our technique.
    For instance, we considered and compared decay processes via maximally entangled and product intermediate states, resulting in the observation of scenarios for quantum correlated and classically correlated loss.
    Also, we modeled dynamical errors via crosstalk-induced bit flips, i.e., a controlled NOT.
    Here, we explored the impact of the initial state, showing that entanglement can significantly alter the process' trajectory and final equilibrium compared to cases where entanglement is suppressed.
    We also showed that our approach is consistent for complex separability-preserving processes even if subsystems of a composite system are randomly exchanged.

    Future applications of our formalism may be concerned with isolating the role of entanglement in quantum thermodynamics.
    For reaching equilibrium, thermalization through ergodicity can be subjected to our tangential projections, for example.
    Furthermore, the impact of entanglement on the Hamiltonian compared to the dissipative part of an open system evolution can be explored in greater detail applying our toolbox.
    Lastly, extending our approach to tangent spaces of other notions of classicality, e.g., via the method from Ref. \cite{SW20}, renders it possible to explore quantum phenomena in open systems beyond entanglement.

    Moreover, from our separability Monte Carlo wave function approach, we formulated for general multipartite system with arbitrary dimension the corresponding master equations that consistently describes the fully separable dynamics.
    Also, we recast those equations in the form of stochastic differential equations.
    A number of properties of these equations were proved, such as consistency with common open-system equations for a single system, previously formulated separability-preserving dynamics if closed systems are considered, and the commonly known master equation when the process under study is already non-entangling.

    Therefore, a generally applicable and easy-to-implement method is put forward to analyze the impact of entanglement within the dynamics of many-body open systems.
    Our results demonstrate a unique insight into the impact of process entanglement even if the initial and final state are not entangled, being specifically tailored for assessing the prospects of noisy intermediate-scale quantum information processing and time-dependent quantum correlations.

\begin{acknowledgments}
    The authors are grateful to Joan Alba for valuable suggestions and comments.
    The authors acknowledge funding through the Ministry of Culture and Science of the State of North Rhine-Westphalia (PhoQC initiative), the Deutsche Forschungsgemeinschaft (DFG, German Research Foundation) via the transregional collaborative research centers TRR 142 (Project C10, Grant No. 231447078), and the QuantERA project QuCABOoSE.
    J.P. acknowledges financial support from the Alexander von Humboldt Foundation (Feodor Lynen Research Fellowship).
\end{acknowledgments}

\appendix

\section{Lindblad master equation for a separable map}
\label{app:Sep}

    In the following, the explicit form of the traditional Lindblad equation is derived for the case of a separable map.
    Consider at time $t$ a system that is in a normalized pure product state $\ket{\psi(t)}=\bigotimes_{k=1}^n \ket{\psi_k(t)}$, which extends to mixtures by sampling over pure states.
    The initial state is propagated to a later time $t+\tau$ by the effective non-Hermitian Hamiltonian $\hat{H}=H-\frac{i}{2}\sum_{a=1}^{m}\big(L^{a}\big)^\dag  L^a$.
    For sufficiently small $\tau$, the evolved state is given by
    \begin{equation}
        \label{eq:NoJump}
        \ket{\psi(t+\tau)}=\big(\mathbbm{1}^{\otimes n}-i\tau \hat{H}\big)\ket{\psi(t)}.
    \end{equation}
    As the evolution is not unitary, the propagated state is not normalized and has a square norm
    \begin{equation}
        \begin{split}
            &\braket{\psi(t+\tau)|\psi(t+\tau)}\\
            =&\bra{\psi(t)}\Big(\mathbbm{1}^{\otimes n}+i \hat{H}^\dag\tau\Big)\Big(\mathbbm{1}^{\otimes n}-i\hat{H}\tau\Big)\ket{\psi(t)}
            =1-q.
        \end{split}
    \end{equation}
    Here, we defined
    \begin{equation}
        q=i\tau\bra{\psi(t)}\big(\hat{H}-\hat{H}^\dag\big)\ket{\psi(t)}=\sum_{a=1}^{m}q_a,\\
    \end{equation}
    with $q_a=\braket{\psi(t)|L^{a\dag} L^a|\psi(t)}$.
    If $q$ is small, no quantum jump occurs, and the normalized state after the time step $\tau$ becomes $\ket{\psi(t+\tau)}/\sqrt{1-q}$.
    For the evolution to be separable, the propagated state has to remain factorized, i.e.,   
    \begin{equation}
        \ket{\psi(t+\tau)}=\bigotimes_{k=1}^n \ket{\psi_k(t+\tau)}.
    \end{equation}
    Comparing with Eq. \eqref{eq:NoJump}, this leads to 
    \begin{equation}
        \begin{split}
        \label{eq:LocOp}
            H&=\sum_{k=1}^{n}\mathbbm{1}^{\otimes(k-1)}\otimes H_k\otimes \mathbbm{1}^{\otimes(n-k)},\\
            \big(L^a\big)^\dag L^a&=\sum_{k=1}^{n}\mathbbm{1}^{\otimes(k-1)}\otimes A_k^a\otimes \mathbbm{1}^{\otimes(n-k)},
        \end{split}
    \end{equation}
    with $A_k^a$ being a Hermitian operator acting on the $k$th subsystem.
    In other words, the nonunitary evolution $\hat{U}={e}^{-i \hat{H}\tau}$ is local \cite{SW17}, indeed.
    
    In cases when a quantum jump occurs, the state collapses,
    \begin{equation}
        \ket{\psi(t+\tau)}=\sqrt{\frac{\tau}{q_a}}L^a\ket{\psi(t)}.
    \end{equation}
    In order for the output to (continuously) evolve into a product state, we must have $L^a=\bigotimes_{k=1}^{n} L^a_k$.
    Together with Eq. \eqref{eq:LocOp}, this yields a condition for separability in terms of the Lindblad operators, viz.
    \begin{equation}
        \label{eq:SepCond}
        \begin{split}
            \big(L^{a}\big)^\dag  L^a&=\bigotimes_{k=1}^{n} \big(L^a_k\big)^\dag L^a_k,\\
            &=\sum_{k=1}^n\mathbbm{1}^{\otimes(k-1)}\otimes A_k^a\otimes \mathbbm{1}^{\otimes(n-k)}.
        \end{split}
    \end{equation}
    For example, if the dynamical map is local, we have $L^a=\mathbbm{1}^{\otimes(a-1)}\otimes L^a_{a}\otimes \mathbbm{1}^{\otimes(n-a+1)}$, for $a=1,\dots,m$ and with $m\leq n$, which satisfies condition \eqref{eq:SepCond}.
    This was to be expected as every local map is separable.

    For an initial product state, at the later time $t+\tau$, the average value of $\rho(t+\tau)$ over the evolution is given by
    \begin{equation}
        \begin{split}
            \rho(t+\tau)=&\ket{\psi(t+\tau)}\bra{\psi(t+\tau)}\\
            &+\tau \sum_{a=1}^mL^a\ket{\psi(t)} \bra{\psi(t)}\big(L^a\big)^{\dag}.\\
        \end{split}
    \end{equation}
    Making use of Eq. \eqref{eq:NoJump}, we can derive an explicit form for the generator of a separable map via
    \begin{equation}
        \rho(t+\tau)=\rho(t)+\tau \mathcal{L}\big(\rho(t)\big).
    \end{equation}
    Dividing the above equation by $\tau$ and taking the limit $\tau\to 0$ eventually yields
    \begin{equation}
        \label{eq:MasterSep}
        \begin{split}
            \frac{{d} \rho}{{d} t}
            =&i\sum_{k=1}^{n}\big[\rho,\mathbbm{1}^{\otimes(k-1)}\otimes H_k\otimes \mathbbm{1}^{\otimes(n-k)}\big]\\
            &-\frac{1}{2}\sum_{a=1}^m\Big\{\mathbbm{1}^{\otimes(k-1)}\otimes  A_k^a\otimes \mathbbm{1}^{\otimes(n-k)},\rho\Big\}\\
            &+\sum_{a=1}^{m}\left(\bigotimes_{k=1}^{n}L^a_k\right)\rho\left( \bigotimes_{k=1}^{n}L^a_k\right)^\dag=\mathcal{L}(\rho).\\
        \end{split}
    \end{equation}
    In contrast to the separability master equation \eqref{eq:SepMaster}, Eq. \eqref{eq:MasterSep} is now deterministic.
    This is due to the generator being linear, i.e., independent of the form of the state $\rho$ as a convex mixture of pure states. 

\section{Consistency for separable maps}
\label{app:Cons}
    
    We are now in a position to show that, for a separable map, both the conventional and the separability master equation yield the identical evolution.
    For a separable map, the above argument ensures the existence of a generator of the form in Eq. \eqref{eq:MasterSep}. Applying Eq. \eqref{eq:RedOp} to Eq. \eqref{eq:LocOp}, as well as to $L^a=\bigotimes_{k=1}^n L_k^a$, yields
    \begin{equation}
    \label{eq:RedSep}
        \begin{split}
            (H)_k=\sum_{i\neq k}^{n} \braket{H_i}\mathbbm{1}+H_k,
            \quad
            \big(L^{a\dag} L^a\big)_k&=\sum_{i\neq k}^{n} \braket{A_i^a}\mathbbm{1}+A_k^a,
            \\
            \text{and}\quad
            (L^a)_k=\prod_{i\neq k}^{n}\braket{L_i^a} L_k^a,\\
        \end{split}
    \end{equation}
    for any $k=1,\dots,n$ and $a=1,\dots,m$.

    With the above consideration, the separability master equation \eqref{eq:SepMaster} for the reduced operators in Eq. \eqref{eq:RedSep} leads to Eq. \eqref{eq:MasterSep}.
    Therefore, it can be concluded that for a separable map the separability master equation is identical to the conventional one.
    In particular, the dependence in Eq. \eqref{eq:SepMaster} on the initial state $\ket{\psi_1\dots \psi_n}$ cancels out, and we are left with a linear equation of motion as sought.

\end{document}